\documentclass[prd, superscriptaddress, tightenlines, longbibliography, nofootinbib, eqsecnum, amsfonts, amsmath, floatfix, twocolumn, notitlepage]{revtex4-2}
\pdfoutput=1

\usepackage[utf8]{inputenc}
\usepackage{mathrsfs}
\usepackage{euscript}
\usepackage{graphics}
\usepackage{graphicx}
\usepackage{amsmath}
\usepackage{amssymb}
\usepackage{gensymb}
\usepackage{bm}
\usepackage[usenames,dvipsnames,svgnames,table]{xcolor}
\definecolor{linkcolor}{rgb}{0.6,0,0}
\definecolor{citecolor}{rgb}{0,0,0.75}
\definecolor{urlcolor}{rgb}{0.12,0.46,0.7}
\usepackage{xspace}
\usepackage{wasysym}
\usepackage{times}
\usepackage{appendix}
\usepackage{comment}
\usepackage{lipsum}
\usepackage[nolist,nohyperlinks]{acronym}
\usepackage{float}
\usepackage{simplewick}
\usepackage{natbib, ifthen}
\usepackage[breaklinks, colorlinks, urlcolor=urlcolor, linkcolor=linkcolor,citecolor=citecolor,pdfencoding=auto]{hyperref}
\usepackage{longtable}
\usepackage{listings}

\providecommand{\planck}{\textit{Planck}}
\providecommand{\Planck}{\planck}

\newcommand{\mksym}[1]{\ifmmode {\rm #1}\else #1\fi}

\providecommand{\Planck}{\textit{Planck}}
\providecommand{\planck}{\Planck}

\providecommand{\text}[1]{\rm{#1}}

\newcommand{\begm}{\begin{pmatrix}}
\newcommand{\enm}{\end{pmatrix}}

\newcommand\ba{\begin{eqnarray}}
\newcommand\ea{\end{eqnarray}}
\newcommand\bea{\begin{eqnarray}}
\newcommand\eea{\end{eqnarray}}

\newcommand\be{\begin{equation}}
\newcommand\ee{\end{equation}}

\newcommand{\boldvec}[1]{{\mbox{\boldmath{$#1$}}}}

\newcommand{\vL}{\boldvec{L}}

\newcommand{\vl}{\boldvec{l}}

\newcommand{\Lobs}{ {L} }
\newcommand{\Lsky}[0]{ {L_{\rm{sky}}} }

\defcitealias{PL2018}{PL2018}

\defcitealias{Carron:2022eyg}{PL4}

\newcommand{\isdraft}[1]{#1}

\newcommand{\JC}[1]{\isdraft{\color{black}{#1}\color{black}}}

\newcommand{\Rmat}[0]{r}
\newcommand{\Amat}[0]{\JC{a}}

\newcommand{\av}[1]{\left \langle #1\right\rangle}

\begin{document}
\title{Real-world CMB lensing quadratic estimator power spectrum response}

\newcommand{\Geneve}{Universit\'e de Gen\`eve, D\'epartement de Physique Th\'eorique et CAP, 24 Quai Ansermet, CH-1211 Gen\`eve 4, Switzerland}

\author{Julien Carron}
\email{julien.carron@unige.ch}
\affiliation{\Geneve}

  \begin{abstract}
I describe a method to estimate response matrices of Cosmic Microwave Background (CMB) lensing power spectra estimators to the true sky power under realistic conditions. Applicable to all lensing reconstruction pipelines based on quadratic estimators (QE), it uses a small number of Gaussian CMB Monte-Carlos and specially designed QE's in order to obtain sufficiently accurate matrices with little computational effort. This method may be used to improve the modelling of CMB lensing band-powers by incorporating at least some of the non-idealities encountered in CMB lensing reconstruction. These non-idealities always include masking, and often inhomogeneous filtering, either in the harmonic domain or pixel space. I obtain these matrices for \planck~latest lensing reconstructions, and then show that the residual couplings induced by masking explain very well the residual multiplicative bias seen on the \planck~simulations, removing the need for an empirical correction.
  \end{abstract}


   \maketitle

\section{Introduction}
The effect of weak gravitational lensing on the Cosmic Microwave Background (CMB) has now been measured to a couple of percent precision, providing a clean probe of the late-time Universe~\citep{Lewis:2006fu}. The relevance of the CMB lensing power spectrum as a cosmological probe is expected to increase further in upcoming years, a major part of its science case being its ability to tightly constrain the neutrino mass scale in combination with other cosmological data sets~\cite{Kaplinghat:2003bh, Abazajian:2016yjj, SimonsObservatory:2018koc}.
Current measurements from the \planck~satellite\cite{PL2018, Carron:2022eyg} and ground-based telescopes\citep{Sherwin:2016tyf, POLARBEAR:2019ywi, Wu:2019hek} use quadratic estimators (QE~\cite{Okamoto:2002ik, Hu:2001kj}) to extract the signal from CMB maps\footnote{\JC{ A recent exception being \cite{Millea:2020iuw} that uses Bayesian techniques to put an approximately $10\%$ constraint on the $\Lambda$CDM lensing power spectrum amplitude $A_\phi$}}. While more efficient methods to extract the spectrum are known~\citep{Hirata:2003ka, Millea:2021had, Legrand:2021qdu}, they will be most useful only when the effective instrumental noise level will be small enough to resolve the lensing-induced polarization $B$-mode over large fractions of the sky, for example with CMB-S4\footnote{\url{https://cmb-s4.org}}\cite{Abazajian:2016yjj} or potentially a high-resolution next-generation space mission such as PICO\cite{NASAPICO:2019thw}.

CMB observations are always only usable on some fraction of the sky.  Estimators designed to work on an idealized, full-sky configuration must account for this in some way, since the presence of the mask will otherwise introduce biases. This also affects their covariance matrices, and eventually will introduce some level of coupling between different multipoles of the CMB power spectra extracted from the data. For the standard CMB spectra, masking effects are relatively straightforward to model and are routinely taken into account\cite[e.g.]{Hivon:2001jp, Chon:2003gx, 2004MNRAS.349..603E}. However, CMB lensing power spectra from quadratic estimators are four-point functions of the data. This renders analytical understanding of the impact of masking, or other non-idealities, considerably more difficult. In practice, corrections obtained from simulations are applied to measured band-powers, as described later on. These corrections are small enough, and the lensing spectrum is smooth enough, that this way of proceeding is believed to be robust enough for the foreseeable future. Nevertheless, a more detailed understanding is certainly desirable. Unfortunately, barely anything quantitative is known on the responses and couplings of CMB lensing estimators away from idealized conditions.
In this short paper I give a method to obtain a good estimate of the measured lensing spectrum response to the true sky spectrum, when extracted with quadratic estimators, at a manageable numerical cost. This makes use of a small number (I will use here 2 for our main results) of noise-free Gaussian CMB simulations to which quadratic estimators designed for this purpose are applied. The method is general enough that it can be used on any quadratic estimator based lensing spectrum extraction pipeline.

I then use this method to obtain the response matrices of the \planck~lensing reconstructions~\cite{PL2018, Carron:2022eyg} for various quadratic estimators, and show that they provide a very good match to the empirical Monte-Carlo correction observed on the complex \planck~NPIPE~\cite{Akrami:2020bpw} simulation suite, that were applied to the published band-powers and likelihoods.  Our rather simple-minded method is based on Eqs.~\eqref{eq:rmatest} and ~\eqref{eq:psi}, which we motivate in Sec.~\ref{sec:mat} after reviewing the relevant elements of CMB lensing quadratic estimators in Sec.~\ref{sec:QE}. An appendix collects additional details on the four-point contractions of the CMB relevant to lensing reconstruction.

\section{Quadratic estimator power spectra}\label{sec:QE}
I follow somewhat closely in this section the notations of the \planck~2015 lensing paper~\cite[Appendix A]{Ade:2015zua}, which provides a fairly exhaustive presentation of standard quadratic estimator theory, and to which I refer for complete expressions and more details. The results obtained here  extend straight-forwardly to the generalized minimum variance estimator of Ref.~\cite{Maniyar:2021msb}, or to the $\kappa$-filtered versions of Ref.~\cite{Mirmelstein:2019sxi}. \JC{Bias-hardened~\cite{Namikawa2013} or `shear-only' estimators~\cite{Schaan:2018tup, Qu:2022qie, Sailer:2022jwt} have weights modified to be more robust to foregrounds, but remain quadratic and can also be treated in the exact same way.}

Lensing introduces statistical anisotropies in the CMB two-point statistics. To linear order, one may write for two CMB fields $X$ and $Z$ the change in covariance due to the lensing potential $\phi$ as
\begin{equation}\begin{split}\label{eq:response}
	\Delta\av{X_{\ell_1 m_1} Z_{\ell_2 m_2}} =\sum_{LM}(-1)^M& \begin{pmatrix} \ell_1 & \ell_2 & L \\ m_1 & m_2 & -M\end{pmatrix} \\& \times\mathcal W^{XZ}_{\ell_1 \ell_2 L}\phi_{LM}
\end{split}
\end{equation} with known covariance response functions $\mathcal W^{XZ}$ given in~\cite{Okamoto:2003zw}.  A quadratic estimator $\bar x[X, Z]$ uses a set of fiducial weights $W^x$ to build
\begin{equation}\begin{split}\label{eq:QE}
\bar x_{LM}	= \frac{(-1)^M} {2} \sum_{\ell_1m_1, \ell_2 m_2} &\begin{pmatrix} \ell_1 & \ell_2 & L \\ m_1 & m_2 & -M\end{pmatrix} \\ &\times W^x_{\ell_1 \ell_2 L} \bar X_{\ell_1 m_1} \bar Z_{\ell_2 m_2},
\end{split}
\end{equation}
where $\bar X, \bar Z$ are inverse-variance weighted $X, Z$ maps, in order to optimize signal to noise by down-weighting noisy pixels or harmonic modes (the specific implementation of this step can vary).
By design, under idealized conditions and full-sky coverage, the estimator \eqref{eq:QE} will then respond diagonally to the lensing potential to linear order according to
\begin{equation}\label{eq:qeresp}
	\av{\bar x_{LM}} = \mathcal R^{x\phi}_L \phi_{LM}
\end{equation} ($\phi_{LM}$ is held fixed in the average). Here $\mathcal R^{x\phi}_L$ sums the QE weights against the responses $\mathcal W^{XZ}$ and the inverse-variance filters mapping $X, Z$ to $\bar X, \bar Z$. An unbiased estimator is then simply obtained inverting the response,
\begin{equation}\label{eq:phi}
	\hat \phi_{LM}^x \equiv \frac{\bar x_{LM}}{\mathcal R^{x\phi}_L}
\end{equation}

Cross-correlating $\hat \phi ^x$ to another estimator built from $\bar y[C, D]$ probes the CMB trispectrum. Defining
\begin{equation}
	C_{L,xy}^{\hat \phi \hat \phi} \equiv \frac{1}{2L + 1} \sum_{M=-L}^L 	\hat \phi_{LM}^x 	\hat \phi_{LM}^{*y},
\end{equation}
and neglecting small complications from the large-scale structure and post-Born bispectra~\cite{Bohm:2016gzt,Beck:2018wud, Bohm:2018omn}, \JC{as well as any contribution from extra-galactic foregrounds}, one has under these idealized conditions~\cite{Hanson:2010rp}
\begin{equation}\label{eq:spec}
	\av{C_{L,xy}^{\hat \phi \hat \phi}} =   C_L^{\phi\phi} + N_{L,xy}^{(0)} + N_{L,xy}^{(1)} 
\end{equation}
where
\begin{itemize}
	\item the first term $C_L^{\phi \phi}$ is sourced by the primary trispectrum contractions, the ones that naturally emerge from Eq.~\eqref{eq:qeresp} involving the CMB sky covariance responses $\mathcal W^{XZ}$ and $ \mathcal W^{CD}$ on each of the QE legs, and each of two legs brings one $\phi$.
	\item $N^{(0)}_L$ is the disconnected 4-point function, proportional to two powers of the data CMB spectra inclusive of the the lensing contribution and instrumental or other noise.
	\item $N_L^{(1)}$ captures the secondary trispectrum contractions, the ones that involve $\mathcal W^{XC}\cdot\mathcal W^{ZD}$ and  $\mathcal W^{XD}\cdot \mathcal W^{ZC}$, the CMB contracting across the two QE legs. These are generally smaller than the primary term except at the highest $L'$s,  but still relevant~\cite{Kesden:2003cc}.
\end{itemize}

On the masked sky, the structure of Eq.~\eqref{eq:spec} remains the same. However the true estimator response of Eq.~\eqref{eq:qeresp} become non-diagonal and position dependent, and is never exactly known. In the lack of a better prescription, one often sticks to the same QE definition\footnote{One also subtracts to $\hat \phi^x$ the average of the QE observed on simulations to remove contributions from anisotropies unrelated to lensing\JC{ (the `mean-field
')}. I stick to Eq.~\eqref{eq:phi} to define $\hat \phi^x$ in this paper.} and normalization, Eqs~\eqref{eq:QE} and~\eqref{eq:phi}. This is a useful approximation which is certainly correct away from the mask boundaries since lensing reconstruction is very local. One crudely accounts for the missing sky area with the rescaling
\begin{equation}\label{eq:fsky}
	C_{L,xy}^{\hat \phi \hat \phi} \rightarrow  \frac{1}{f_{\rm sky}}	C_{L,xy}^{\hat \phi \hat \phi}
\end{equation}
where $f_{\rm sky}$ is the unmasked sky fraction (or the mean fourth power of the mask if a non-boolean mask is used, as proposed by Ref.~\cite{Benoit-Levy:2013uxb}).

All of the terms in Eq.~\eqref{eq:spec} respond to the sky lensing spectrum. However, $N^{(0)}_L$ can be removed accurately by using QE's built from a mix of data and simulations~\citep{Namikawa2013, Ade:2013mta, Story:2014hni}. Unlike an analytical $N^{(0)}$, this bias estimate built from QE's properly contains all masked-induced couplings, and its accuracy is not degraded by the approximate normalization since it is acting on the unnormalized estimators $\bar x$ and $\bar y$. Within the fiducial cosmological model, the same holds for $N_L^{(1)}$\citep{Story:2014hni}, which can also be evaluated by QE's on pairs of simulations tuned to capture the secondary contractions. The construction of these accurate bias estimates is reviewed in the appendix.

\newcommand{\Nzro}[0]{\hat N^{(0)}}
\newcommand{\None}[0]{\hat N^{(1)}}
For these reasons, when averaged over simulations with consistent cosmology, the subtraction of the lensing biases is correct using specially designed biases estimates (`MC-$\Nzro$' and `MC-$\None$'), but the remaining primary term will be somewhat off. One can use this to provide a definition of a spectrum response matrix that captures the primary trispectrum contractions only:

\begin{equation}\label{eq:rmat}
\begin{split}
	\av{C_{L,xy}^{\hat \phi \hat \phi} - \text{MC-}\Nzro_L-  \text{MC-}\None_L}_{\rm MC}  \equiv \sum_{\Lsky}  \Rmat^{xy}_{L\Lsky} C_{\Lsky}^{\phi \phi}
\end{split}
\end{equation}
Within $\Lambda  $CDM (and probably for most other models) the lensing spectrum is smooth and largely featureless. This results in a multiplicative bias affecting the reconstructed spectrum at multipole $L$. The factor $f_{\rm sky}$ in Eq~\ref{eq:fsky} has been included into this definition of the matrix, so that we expect the bias to be smaller then $f_{\rm sky}$.

\section{Building the spectrum response matrices}\label{sec:mat}

\newcommand{\pLsky}[0]{\hat \phi^{x_{\Lsky}}}
\newcommand{\pyLsky}[0]{\hat \phi^{y_{\Lsky}}}
\newcommand{\pG}[0]{\hat \phi^{x_{g}}}
\newcommand{\pGy}[0]{\hat \phi^{y_{g}}}

To isolate the spectrum response to a sky harmonic multipole $L_{\rm sky}$, one may imagine building simulated lensed CMB skies, $X^{\Lsky}, Z^{\Lsky}$, where the lensing deflection field was populated with modes at multipole $\Lsky$ only, and building the quadratic estimator
\begin{equation}
	 \hat \phi_{LM}[X^{\Lsky}, Z^{\Lsky}]\equiv \pLsky_{LM}
\end{equation} on this simulation.  Its spectrum then traces the entire column of the response matrix $r^{xy}_{L \Lsky}$. While there is nothing fundamentally wrong with this simple idea, it suffers in practice from a little series a defects:
\begin{enumerate}
	\item CMB fluctuations generically dominate the lensing reconstruction noise budget. Hence, even using simulations free of instrumental noise, $\pLsky$ will have (depending on multipole, sometimes considerable) Gaussian lensing noise $N^{(0)}$. While this may be removed, its sample variance would remain, requiring a very large number of simulations to beat it down.
	\item \JC{A similar } issue holds on the largest scales where $\pLsky$ is typically dominated by anisotropies unrelated to lensing, collectively called the `mean-field'. \JC{When naively estimated averaging reconstructions from a number of  independent simulations, subtraction of the mean-field introduces Monte-Carlo noise proportional to $N^{(0)}$.}
	\item Non-perturbative effects, relevant to lensing reconstruction, are essentially absent from the spectrum of $\pLsky$, since all other multipoles have been set to zero. 
\end{enumerate}
Further, the spectrum of $\pLsky$ would contain both primary and secondary trispectrum contractions which does not match our definition \eqref{eq:rmat}( though this should not be considered as an issue in general, as this would certainly also be an interesting matrix to look at, if feasible).

The first two points induce unacceptable additional numerical cost if treated too naively. Cancelling reconstruction noise and mean-field at the map-level seems the most natural and efficient way forward. Hence, with $X^{g}, Z^{g}$ the Gaussian unlensed CMBs of the same simulation, consider the signal-free lensing QE $ \hat \phi[X^{g},Z^{g}] \equiv \pG$  and the difference 
\begin{equation}\label{eq:phidiff}
 \pLsky_{LM} - \pG_{LM}. 
\end{equation}
The $N^{(0)}$ reconstruction noise of the spectrum of this map now scales with the difference in power between $\Lsky$-only lensed and unlensed rather than with the absolute lensed power, which is smaller by orders of magnitude. Likewise the mean-field is subtracted to very high accuracy. Building this difference also affects the signal: in the combination\JC{
\begin{equation*}
\begin{split}
&\frac{1}{(2L + 1)} \sum_{M = -L}^L \left(\pLsky_{LM} - \pG_{LM}\right)\left(\pyLsky_{LM} - \pGy_{LM}\right)^\dagger\\
& =C_{L,x_{\Lsky}y_{\Lsky}}^{\hat\phi\hat\phi} -	C_{L,x_{\Lsky}y_{g}}^{\hat\phi\hat\phi}- C_{L,x_{g}y_{\Lsky}}^{\hat\phi\hat\phi} +  C^{\hat\phi\hat\phi}_{L,x_{g}y_{g}}
\end{split}
\end{equation*}}
contractions of the $N^{(1)}$ type are present in the first three terms (see the discussion in the appendix), while the primary signal is unaffected since only the first term gives a contribution. 

Among the possibilities to address point \JC{3}, I proceed in the following approximate manner, which has the merits of operational simplicity and of removing the need to deflect any maps: non-perturbatively, it is well known~\cite{Hanson:2010rp} that usage of the lensed CMB spectra instead of the unlensed spectra provide an accurate match to the lensing response (in temperature and for low noise levels, an even better better prescription is that of the grad-lensed spectra~\cite{Fabbian:2019tik}, but I do not aim here to that level of precision). Usage of the lensed spectra in the quadratic estimator weights is also known to be both more optimal and to remove terms beyond linear in the spectra ($N^{(2)}$ lensing bias~\cite{Hanson:2010rp}). For these reasons,
I consider Gaussian unlensed CMB's but possessing lensed rather than unlensed spectra, and expand perturbatively the action of the deflection. Writing schematically~\cite{Challinor:2002cd} the leading term of the deflected Stokes maps as
\begin{equation*}
	X^{\Lsky}(\hat n) \sim X^g(\hat n) + \alpha^{\Lsky}(\hat n)\cdot \partial X^g(\hat n),
\end{equation*}
where $\alpha^{\Lsky}$ is the deflection vector populated with modes at multipole $\Lsky$ only,
and plugging into~\eqref{eq:phidiff} motivates the following quadratic estimators\JC{ (see also the appendix)}
\begin{equation}\label{eq:psi}
	\hat \psi^{x_\Lsky} \equiv \hat \phi [X^g,  \alpha^{\Lsky}\cdot \partial Z^g ] + \hat \phi [\alpha^{\Lsky}\cdot \partial X^g , Z^g]
\end{equation}
I found that the spectra and cross-spectra of these maps provide with a single Monte-Carlo, in the configurations that I tested, an estimate of the row $\Rmat^{xy}_{\Lobs \Lsky}$ with good relative accuracy for each entry. At the highest-$\Lobs$'s, the residual $N_L^{(0)}$ noise can still dominate, and a subset of the $N^{(1)}$-type contractions are still present. A simple way to get rid of these biases is then simply to build two independent simulations $C^{'g}, D^{'g}$ but use the same deflection and take a cross-spectrum. This cross-spectrum gives then an unbiased estimate of the coupling matrix defined in Eq.~\eqref{eq:rmat},
\begin{equation}\label{eq:rmatest}
	\hat r^{xy}_{\Lobs \Lsky}  = \frac{1}{C_{\Lsky}^{\phi\phi}} \frac{1}{(2L + 1) f_{\rm sky}}\sum_{M=-L}^L\hat \psi_{LM}^{x_\Lsky}\cdot	 \hat \psi^{y'_\Lsky *}_{LM}.
\end{equation}
The Monte-Carlo noise in the estimates is to a large extent due to the cosmic variance of the lenses. This can be further reduced by considering the ratio to the realized $C_\Lsky^{\phi\phi}$.

A complete estimate of the response matrix can thus be filled by calculating $L_{\rm sky, \max}$ quadratic estimators starting from a few Gaussian maps. For a lensing pipeline similar to the~\planck~one, the bulk of the numerical work goes into the inverse-variance filtering of each of the $\alpha^\Lsky$-weighted maps. This remains non-negligible in absolute terms, but is a very small addition to the the cost of a full lensing reconstruction on  data.

\JC{The lensing spectrum is very red and in practice it is convenient to work with the rescaled matrix
\begin{equation}\label{eq:amat}
	\hat a^{xy}_{\Lobs \Lsky} \equiv \frac{1}{C_{L}^{\phi\phi}}\hat r^{xy}_{\Lobs \Lsky} C_{\Lsky}^{\phi\phi}.
\end{equation}
which linearly relates the amplitudes $A^{\phi}_{\Lsky}$ of the true spectrum in units of $C_{\Lsky}^{\phi\phi}$ to the observed ones $\hat A^\phi_L$.}
\subsection{$\Planck$-lensing coupling matrix}
\begin{figure}
   \centering
     \includegraphics[width=\hsize]{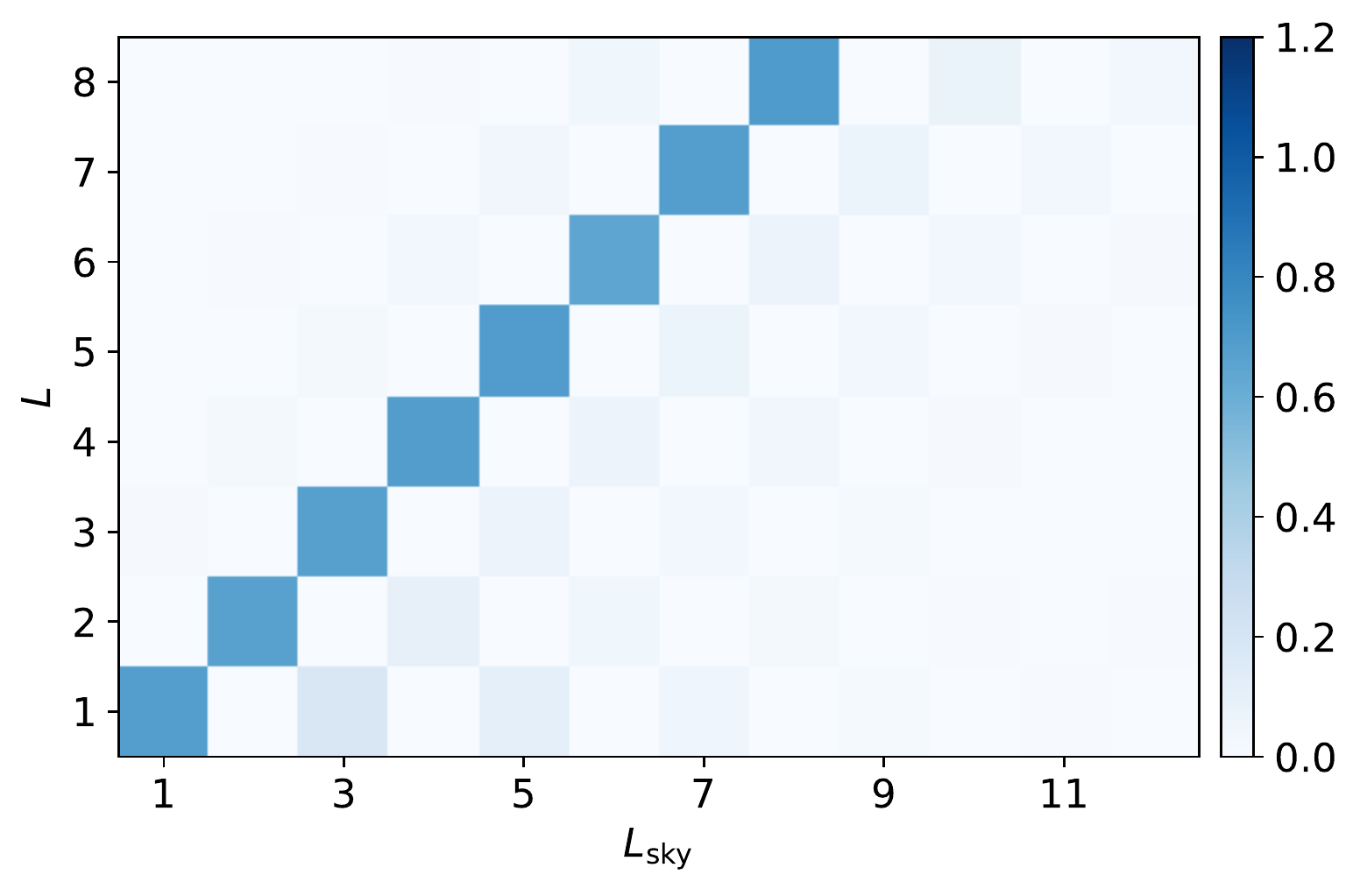}
      \caption{Response matrix of the minimum variance \planck~PR4 2018-like lensing estimator at low lensing multipoles, Eq.~\ref{eq:rmat}\JC{ , normalized following \eqref{eq:amat}}. This was obtained using 12 Monte-Carlos. At higher multipoles, the matrix displays an almost exact symmetric structure with constant diagonals, shown on Fig.~\ref{fig:rows}.}
         \label{fig:lowL}
 \end{figure}
 \begin{figure}
   \centering
     \includegraphics[width=\hsize]{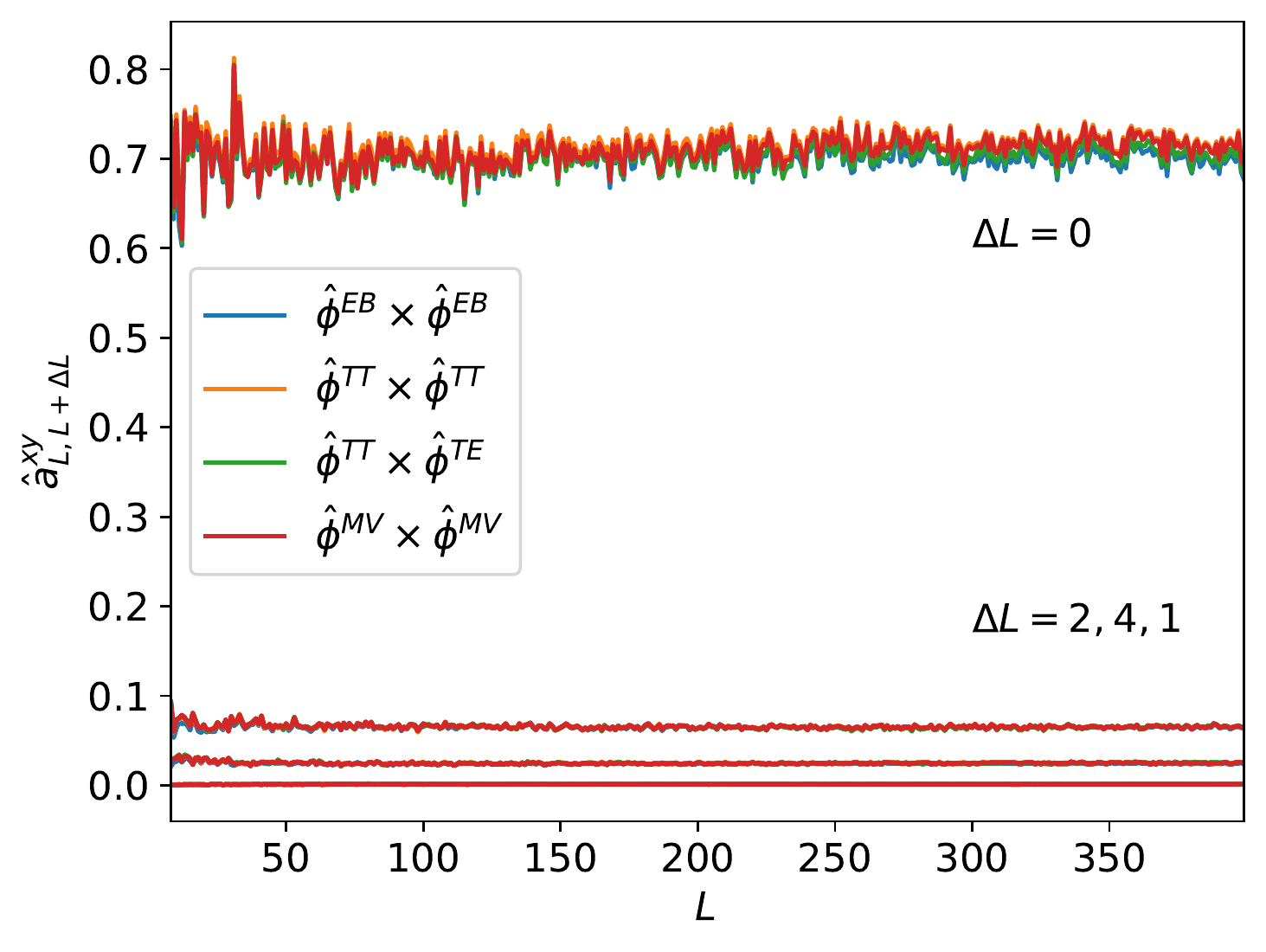}
      \caption{Lensing spectrum response matrix for \planck~lensing spectrum reconstructions. Shown are $\hat \Amat^{xy}_{L,L+\Delta L} $ for $\Delta L = 0, 2, 4$ and $1$ from top to bottom (the matrix is very close to symmetric with constant diagonals.) Displayed are results for different estimators $x$ and $y$ as indicated in the legend, but curves are hardly distinguishable from each other.\JC{ For accurate calculations, it is necessary to include additional rows, which are all very small individually but this is compensated by a not particularly fast power-law like decay with $\Delta L$.}}
         \label{fig:rows}
 \end{figure}
 \begin{figure}
   \centering
   \includegraphics[width=\hsize]{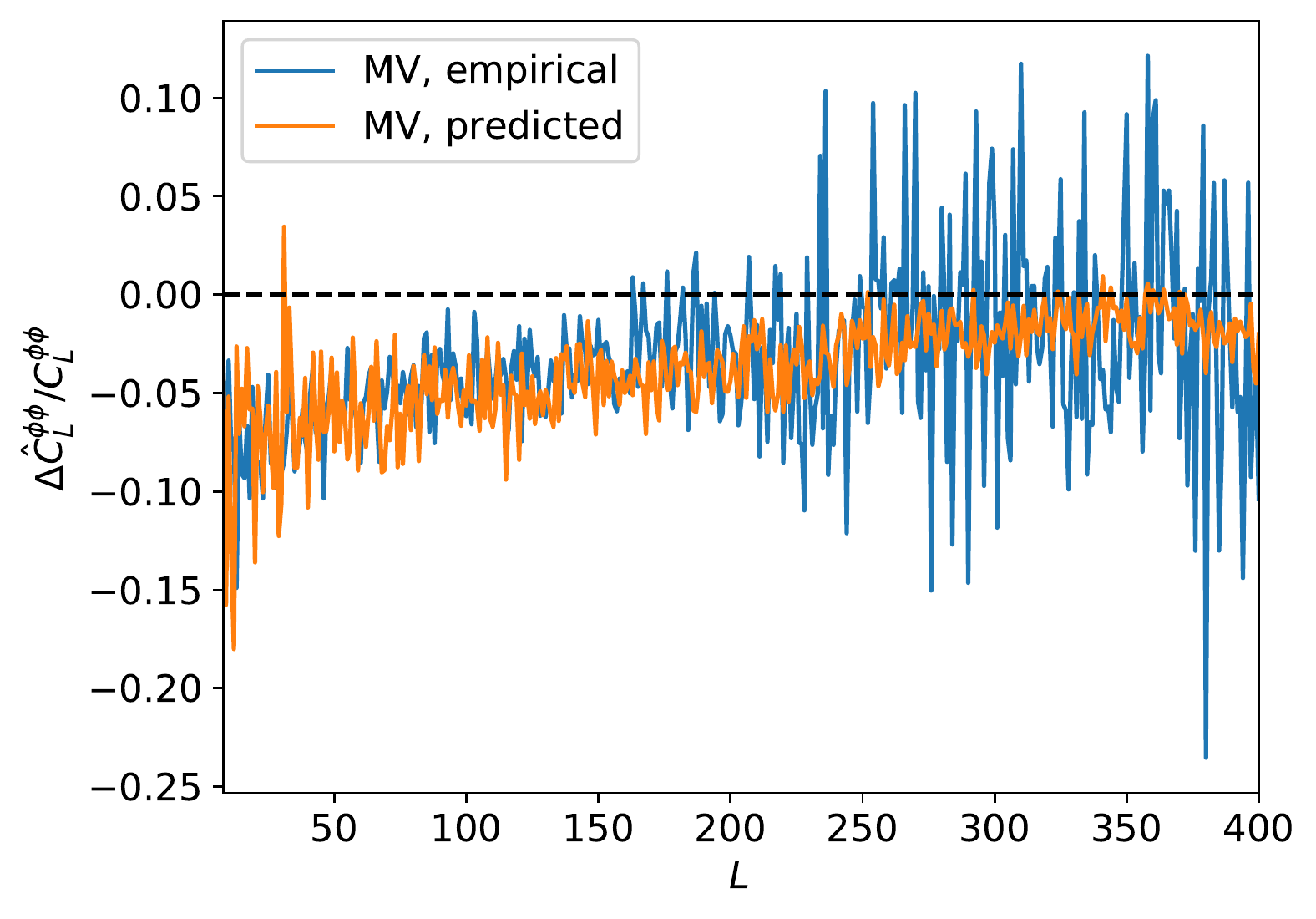}
      \caption{Unbinned empirical \planck~lensing MV reconstruction Monte-Carlo correction (as defined by Eq.\eqref{eq:bamc}, blue), and the prediction obtained from Eq.~\eqref{eq:rmatest} in this work (orange), using 2 simulations per each sky mode, on the $L$-range $8\leq L \leq 400$ used for the lensing likelihood. The blue points were obtained averaging 480 $\planck$ NPIPE simulations, with substantial Monte-Carlo noise remaining at high $L$.}
         \label{fig:bamc}
 \end{figure}
I tested this method on $\Planck$ latest lensing reconstructions from Ref.~\cite{Carron:2022eyg}, that use the most recent NPIPE \planck~CMB maps.  The version of the pipeline using homogeneously filtered maps is virtually identical to the previous \Planck~lensing likelihood 2015 and 2018 releases~\cite{Ade:2015zua, PL2018} so that I expect only very minor differences. The CMB-filtering in this pipeline proceeds by inverse total variance weighting the CMB according to an approximate CMB likelihood model, using an isotropic beam and homogeneous noise model outside of the lensing mask, keeping $67\%$ of the sky. This step is quite efficient at undoing the mask-induced mode-coupling at the level of the CMB maps, and is performed by conjugate-gradient inversion. I obtain then the response matrix from Eq.~\eqref{eq:rmatest}, using two independent estimates in order to get a very minimal handle on the Monte-Carlo noise of the estimate, and I have used a few more ($\sim 10$) simulations at the lowest lensing multipoles $L \leq 8$. The Monte-Carlo noise is predominantly sourced by the cosmic variance of the input lensing potential spectrum, and therefore is higher there. 

The low-$L$ part of the matrix is displayed on Fig.~\ref{fig:lowL}. A chessboard-alike pattern is clearly visible, with elements suppressed when $|\Delta L| = |\Lobs - \Lsky|$ is an odd number. This is because of the approximate symmetry of the \planck~lensing mask with respect to the galactic equator, resulting in enhanced even mask multipoles in galactic coordinates. At higher $L$, the response matrix appears very close to symmetric, and with constant diagonals. It is also very much the same for different types of quadratic estimators. The main and first few offset diagonals are shown on Fig.~\ref{fig:rows}, for three of the most statistically powerful \planck~estimators from bottom to top (MV$\times$MV, TT$\times$TT and TT$\times$TE). Also shown is the physically different EB$\times$EB, which unlike temperature, is insensitive the large-scale convergence modes (and is almost pure noise at these noise levels). All estimators were obtained using the public \planck~lensing pipeline \texttt{plancklens}\footnote{\url{https://github.com/carronj/plancklens}}. The different estimators can hardly be distinguished from each other, however. The size of the couplings are about 9\% and 3.5\% of the main diagonal for the most relevant entries $\Delta L = 2$ and $4$.

Using 480 NPIPE simulations and their lensing reconstructions, I build the difference of the spectrum estimate to the simulation input,
\begin{equation*}\label{eq:bamc}
\Delta \hat C^{ \phi \phi}_{L, xy} \equiv	\av{C_{L,xy}^{\hat \phi\hat \phi} - \text{MC-}\hat N_{L,xy}^{(0)} - \text{MC-}\hat N_{L,xy}^{(1)} - C_L^{\phi\phi}}_{\text{MC's}}.
\end{equation*}
In this equation, the MC-$\hat N^{(0)}$ and MC-$\hat N^{(1)}$ biases are computed from simulations exactly as described in Ref.~\cite{PL2018}. Figure~\ref{fig:bamc} shows this mismatch in blue for the MV reconstruction, which reaches $5-10\%$ at the lowest lensing multipoles. Displayed is the range $8 \leq L \leq 400$ which is the one used for the \planck~lensing likelihoods. The prediction from this work is
\begin{equation}
\frac{\Delta \hat C^{\phi\phi}_{L,xy}}{C_L^{\phi\phi} } =	\sum_{\Lsky} \left( \hat \Amat^{xy}_{L \Lsky} - \delta_{L \Lsky} \right).
\end{equation}
This is shown in orange, where again I have used a pair of simulations to estimate the matrix. While both curves suffers from some level of Monte-Carlo noise, the prediction clearly provides a good match to the empirical behavior. \JC{I note that while individual coefficients are small in absolute terms after $\Delta L = 4$, they decay relatively slowly (power-law) and it is necessary for a good relative accuracy on the correction to includes terms beyond that (up to a few tens of multipoles on the highest multipoles shown there).}

\JC{Modelling the bandpowers using the coupling matrix provides an alternative modelling choice to that of applying a multiplicative correction as done so far (both methods with some levels of Monte-Carlo noise). Using the binning procedure of the \planck~analyses, we find that the predictions of the spectrum amplitude at the fiducial cosmology differ by only 0.3\%, almost a factor of 10 smaller than the \planck~constraint.}

\section{Conclusions}
I gave a simple prescription to obtain the response matrix sourced by masking or other non-idealities for lensing power spectrum reconstruction from CMB data, and demonstrated it on \planck~maps. This simulation-based method captures the primary lensing trispectrum contractions with the help of very few simulations, and a set of tuned quadratic estimators. I obtained the matrix for the \planck~lensing reconstructions and found that they appear essentially independent of the estimator type, with a simple symmetric structure with constant diagonals, similar to that describing the inverse-variance filtered CMB spectra. \JC{The method automatically includes the combined signatures of the non-idealities present in the filtering, for example those of point-source, cluster and galactic masking, which can impact in a sensibly different manner the band-powers~\cite{Wu:2019hek}.}

This method could be improved further. For example, the treatment of non-perturbative effects is only approximate. One could certainly take these better into account by working with deflected instead of Gaussian CMB's, and extracting with difference maps in a way similar to that presented here the response to a sky lensing mode\footnote{I thank Antony Lewis for this suggestion.}. Working with deflected maps comes with some additional numerical cost, but in recent years this has become easily manageable. Since the structure of the response matrix is so smooth, it is also possible, rather trivially, to use a lot less computational power than done here simply by sampling in $\Lsky$ and interpolating.

 I showed explicitly that the \planck~Monte-Carlo corrections applied to the published bandpowers are consistent with masking-induced residual mode-coupling, and that this mode-coupling is at most a few percent level or below everywhere with the only exception of $\Delta L = 2$. This confirms quite explicitly the recent practice of applying this correction multiplicatively and not additively as was the standard for the first measurements of the lensing spectrum. Given the moderate size of this correction, and the \JC{smoothness of the lensing potential power spectrum}, this choice was largely unconsequential anyways, with main effect a slight underestimation of the covariance matrix at low-$L$. Alternatively, using this matrix to forward-model the cosmological model prediction would remove the need for a Monte-Carlo correction altogether, or in a more general situation allow a clean separation from the effect of masking and filtering from other potential contaminants.

The methods of this paper can be applied to any quadratic estimator pipeline, hence could be useful for other data sets, in particular from the ground, where the reduced sky coverage induces larger mode-couplings, and where unlike \planck~anisotropic filtering in the harmonic domain is often mandatory.

\begin{acknowledgements}
I thank Antony Lewis and Louis Legrand for comments on the draft and discussions. I also thank Anthony Challinor for pointing out an error in the first version of the paper and other comments that lead to improvements. I also thank the anonymous referee for a most careful and useful report. I acknowledge support from a SNSF Eccellenza Professorial Fellowship (No. 186879). This research used resources of the National Energy Research Scientific Computing Center (NERSC), a U.S. Department of Energy Office of Science User Facility located at Lawrence Berkeley National Laboratory.
\end{acknowledgements}

\appendix
\onecolumngrid
\section{Lensing four-point contractions}
\newcommand{\W}[0]{ {\mathcal W} }
In this appendix we discuss for completeness the construction of $\text{MC-}\hat N_{L,xy}^{(0)}$ and $\text{MC-}\hat N_{L,xy}^{(1)}$ lensing biases in non-ideal conditions, and some details pertaining to the main text. 
In practice, the inverse-variance weighted maps $\bar X, \bar Z$ that are fed into the estimators \eqref{eq:QE} are related to the sky $X$ and $Z$ modes by some process which can be quite complicated in many ways but still essentially linear. The expectation of the auto- and cross-spectra between QE's cannot be predicted analytically exactly anymore, but still probes the sky four-point function. As always one splits this four-point function as the sum of a Gaussian-like, disconnected piece, which is considered essentially reconstruction noise, and a non-Gaussian part, connected contribution, which is essentially the sought-after signal.

For simplicity, we use now flat-sky notation. Defining $\vL^{\rm sky} = \vl_1 + \vl_2$ and $\vL'^{\rm sky} = \vl'_1 + \vl'_2$, we may write the disconnected part of the general four-point function
\begin{align}\label{eq:disconnected}
\av{X_{\vl_1} Z_{\vl_2} C_{\vl'_1}D_{\vl'_2}}_{\rm disconn.} = &(2\pi)^2 \delta^{D}(\vL^{\rm sky}) (2\pi)^2 \delta^{D}(\vL'^{\rm sky}) C_{\vl_1}^{XZ}C_{\vl'_1}^{CD} &\textrm{(subtracted by the mean-field)} \\
+&(2\pi)^2 \delta^{D}(\vl_1 + \vl'_1) (2\pi)^2 \delta^{D}(\vl_2 + \vl'_2) C_{\vl_1}^{XC}C_{\vl_2}^{ZD} &\textrm{   $\left(\in N_{L, xy}^{(0)}\right)$}\label{eq:n01}\\
+&(2\pi)^2 \delta^{D}(\vl_1 + \vl'_2) (2\pi)^2 \delta^{D}(\vl_2 + \vl'_1) C_{\vl_1}^{XD}C_{\vl_2}^{ZC} &\textrm{   $\left(\in N_{L, xy}^{(0)}\right)$}\label{eq:n02}
\end{align}
The spectra here all include their instrumental noise as well as lensing contribution. Under idealized conditions, the lensing estimator multipole $L$ obeys $L = L^{\rm sky}$ so that the first term only affects the irrelevant lensing estimator monopole. In the real world, unavoidable post-filtering residual couplings result in a relevant contribution on large-scales, which is removed by considering the mean-field subtracted QE $\hat\phi^{x} - \av{\hat \phi^x}$, where the average is performed on the available simulation set. Let now primed maps ($X'$, etc) denote maps of the same type but statistically independent from the original (non-primed) maps. Then it is apparent that
\begin{equation}\label{eq:mcn0}
\av{	\hat \phi[X, Z'] \cdot  	\hat \phi[C, D']+ 	\hat \phi[X, Z'] \cdot  	\hat \phi[C', D]} \equiv \text{MC-}\hat N_{L,xy}^{(0)}
\end{equation}
captures \eqref{eq:n01} and \eqref{eq:n02} in the first and second term respectively, inclusive of the complications sourced by the non-trivial filtering relating $\bar X$ to $X$ etc. The $\cdot$ stands here for the cross-spectrum.
The lensing maps in this equation require no mean-field subtraction, since term \eqref{eq:disconnected} does not enter \eqref{eq:mcn0} by construction. 

Let us now consider the non-Gaussian part of the four-point function \cite{Kesden:2003cc, Hanson:2010rp}. Perturbatively, a simple way to collect all the leading terms is to list the connected pairings of $(X + \alpha \cdot \partial X)(Z + \alpha \cdot \partial Z)(C + \alpha \cdot \partial C)(D + \alpha \cdot \partial D)$ to first order in $C^{\phi\phi}$. In this work we use variations of the four-point function were some maps are lensed and other Gaussian. In order to make notation more transparent to this, we indicate in the response functions $\W$ defined in \eqref{eq:response} which map was lensed with a tilde. For example $\alpha \cdot \W^{\tilde X Z}$ is sourced perturbatively by the contraction of $\alpha \cdot \partial X$ with $Z$. The full response function of the lensed CMB in \eqref{eq:response} is $\W^{\tilde X Z} +\W^{X \tilde Z} $. Strictly speaking, working perturbatively results in usage of the unlensed spectra in the responses. It is well known that usage of the lensed or grad-lensed spectra resums many of the non-perturbative terms giving more accurate results. The general form of the connected contribution to first order is
\begin{align}\label{eq:trisp}
\av{X_{\vl_1} Z_{\vl_2} C_{\vl'_1}D_{\vl'_2}}_{\rm conn.} = &(2\pi)^2 \delta^D(\vL^{\rm sky} + \vL'^{\rm sky}) C^{\phi\phi}_{\vL^{\rm sky}  }~\left( \W^{\tilde X Z}_{\vl_1\vl_2} +  \W^{X \tilde Z}_{\vl_1\vl_2} \right)\left(\W^{\tilde C D}_{\vl'_1\vl'_2} +  \W^{C \tilde D}_{\vl'_1\vl'_2} \right)&\textrm{ (primary contr.)}  \\ \label{eq:trisp1}
+ & (2\pi)^2 \delta^D(\vL^{\rm sky} + \vL'^{\rm sky})  C^{\phi\phi}_{\vl_1 + \vl_2'} \left(\W^{\tilde X D}_{\vl_1\vl'_2}  + \W_{\vl_1 \vl_2'}^{X \tilde D}\right) \left(\W^{\tilde Z C}_{\vl_2\vl'_1}  + \W_{\vl_2 \vl_1'}^{Z \tilde C}\right)&\textrm{ (secondary, $\in N_{L, xy}^{(1)}$)} \\ 
+&(2\pi)^2 \delta^D(\vL^{\rm sky} + \vL'^{\rm sky}) C^{\phi\phi}_{\vl_1 + \vl_1'} \left(\W^{\tilde X C}_{\vl_1\vl'_1}  + \W_{\vl_1 \vl_1'}^{X \tilde C}\right)\left(\W^{\tilde Z D}_{\vl_2\vl'_2}  + \W_{\vl_2 \vl_2'}^{Z \tilde D}\right) &\textrm{ (secondary, $\in N_{L, xy}^{(1)}$)} \label{eq:trisp2}
\end{align}
Say the maps $X'_\phi$ etc have as before independent unlensed CMB to that of $X$, but share the same lensing deflection field. Crossing QE's just as for $\text{MC-}\hat N_{L,xy}^{(0)}$ will suppress in the same way the primary contractions, but the secondaries \eqref{eq:trisp1} and  \eqref{eq:trisp2} remain intact. This way one gets
\begin{equation}\label{eq:mcn1}
\av{	\hat \phi[X_\phi, Z'_\phi] \cdot  	\hat \phi[C_\phi, D'_\phi]+ 	\hat \phi[X_\phi, Z'_\phi] \cdot  	\hat \phi[C'_\phi, D_\phi]} \equiv \text{MC-}\hat N_{L,xy}^{(1)} + \text{MC-}\hat N_{L,xy}^{(0)}  
\end{equation}
again inclusive of the effect of the non-ideal filters applied to the maps. Combining with \eqref{eq:mcn0} gives $\text{MC-}\hat N_{L,xy}^{(1)}$. One may use CMB-only maps in \eqref{eq:mcn1} to reduce somewhat the Monte-Carlo noise. 

In the main text Eq.\eqref{eq:phidiff} we introduced QEs of the type $\hat \phi^x - \hat \phi^{x_g}$, the difference of a QE built on lensed maps to that built on a Gaussian version of the same maps. After taking a cross-spectrum $(\hat \phi^x - \hat \phi^{x_g}) \cdot (\hat \phi^y - \hat \phi^{y_g}) $, the primary contractions are the same than those of $\hat \phi^{x}\cdot\hat \phi^{y}$. However, the secondaries where the lensing map contracts on the same QE are suppressed by the cross-spectra $-\hat \phi^x \cdot \hat \phi^{y_g}$ and $-\hat \phi^{x_g} \cdot \hat \phi^{y}$. These are the terms $\propto \W^{\tilde X C}\W^{\tilde Z D}$ and $ \W^{X \tilde C}\W^{Z \tilde D}$ in \eqref{eq:trisp2}, and similarly in \eqref{eq:trisp1}. Those secondaries with the lensing map contracting across QE's remain (the terms $\propto \W^{X \tilde C}\W^{\tilde Z D}$ and $\W^{\tilde X C}\W^{Z \tilde D}$ in \eqref{eq:trisp2} and similarly in \eqref{eq:trisp1} ). All of these remaining secondaries, together with the residual $N^{(0)}_{L, xy}$, are suppressed in \eqref{eq:psi}, with usage of independent CMB's in the second QE.

In this work we isolated the primary contractions, and with massively reduced $N^{(0)}$ bias and MC-noise. One can do the same for any other particular contraction. For instance,
\begin{equation}
		\hat \phi[\alpha \cdot \partial  X^g,  \alpha \cdot\partial  Z'^g]  \cdot \hat \phi[C^g, D'^g] 
		\end{equation}
gets as only connected contribution the term $\propto \W^{\tilde X C} \W^{\tilde ZD}$ in \eqref{eq:trisp2}, with no $N^{(0)}$ noise. Tuning the spectra of the Gaussian maps to the non-perturbative response spectra be used again to account for non-perturbative effects. This generalizes easily to the other terms. For example, a temperature-only, low $N^{(0)}$, $N^{(1)}$-estimate is provided by
\begin{equation}
	\av{4\:\hat \phi[\alpha \cdot \partial T^g, \alpha \cdot \partial T'^g] \cdot \hat \phi[T^g, T'^g] + 4\:\hat \phi[T^g, \alpha \cdot \partial T'^g] \cdot \hat \phi[\alpha \cdot \partial T^g, T'^g]}.
\end{equation}
\bibliography{lensingbib, texbase/cosmomc,texbase/antony}

\end{document}